\documentclass{elsart}

\usepackage{graphicx}

\usepackage{amssymb}

\begin{document}

\newcommand{\newc}{\newcommand}

\newc{\be}{\begin{equation}}
\newc{\ee}{\end{equation}}
\newc{\ba}{\begin{eqnarray}}
\newc{\ea}{\end{eqnarray}}
\newc{\bea}{\begin{eqnarray*}}
\newc{\eea}{\end{eqnarray*}}
\newc{\D}{\partial}
\newc{\ie}{{\it i.e.} }
\newc{\eg}{{\it e.g.} }
\newc{\etc}{{\it etc.} }
\renewcommand{\etal}{{\it et al.}}
\newcommand{\nn}{\nonumber}

\newc{\ra}{\rightarrow}
\newc{\lra}{\leftrightarrow}
\newc{\lsim}{\buildrel{<}\over{\sim}}
\newc{\gsim}{\buildrel{>}\over{\sim}}

\begin{frontmatter}

% Title, authors and addresses

% use the thanksref command within \title, \author or \address for footnotes;
% use the corauthref command within \author for corresponding author footnotes;
% use the ead command for the email address,
% and the form \ead[url] for the home page:
% \title{Title\thanksref{label1}}
% \thanks[label1]{}
% \author{Name\corauthref{cor1}\thanksref{label2}}
% \ead{email address}
% \ead[url]{home page}
% \thanks[label2]{}
% \corauth[cor1]{}
% \address{Address\thanksref{label3}}
% \thanks[label3]{}

\title{Exponential Cardassian Universe}
\author{Dao-jun Liu},
\author{Chang-bo Sun}, \author{Xin-zhou Li} \ead{kychz@shnu.edu.cn}
\address{Shanghai United Center for Astrophysics(SUCA),\\
 Shanghai Normal University, 100 Guilin Road, Shanghai 200234,China}

% use optional labels to link authors explicitly to addresses:
% \author[label1,label2]{}
% \address[label1]{}
% \address[label2]{}

\begin{abstract}
% Text of abstract
The expectation of explaining cosmological observations without
requiring new energy sources is forsooth worthy of investigation. In
this letter, a new kind of Cardassian models, called exponential
Cardassian models, for the late-time universe are investigated in
the context of the spatially flat FRW universe scenario. We fit the
exponential Cardassian models to current type Ia supernovae data and
find they are consistent with the observations. Furthermore, we
point out that the equation-of-state parameter for the effective
dark fluid component in  exponential Cardassian models can naturally
cross the cosmological constant divide $w=-1$ that observations
favor mildly without introducing exotic material that destroy the
weak energy condition.
\end{abstract}

\begin{keyword}
% keywords here, in the form: keyword \sep keyword

% PACS codes here, in the form: \PACS code \sep code
\PACS  98.80.-k
\end{keyword}
\end{frontmatter}

\section{Introduction}
The current accelerating expansion of the universe indicated by the
astronomical measurements from Supernovae type Ia (SNeIa)
\cite{SNeIa} (see \cite{SnST,SnLS} for most recent results) as well
as accordance with other observations such as the cosmic microwave
background (CMB) \cite{WMAP} and galaxy power spectra \cite{SDSS}
becomes one of the biggest puzzles in the research field of
cosmology. There are lots of approaches to unriddle this puzzle. One
popular theoretical explanation approach is to assume that there
exists a mysterious energy component, dubbed dark energy, with
negative pressure, or equation of state with $w=p/\rho < 0$ that
currently dominates the dynamics of the universe (for a review see
\cite{DEReivew}).  Such a component makes up $70\%$ of the energy
density of the universe yet remains elusive in the context of
general relativity and the standard model of particle physics. In
recent years, many candidates for dark energy have been explored.
Besides a cosmological constant, one popular candidate source of
this missing energy component is a slowly evolving and spatially
homogeneous scalar field, referred to as
 "quintessence" with $w>-1$ \cite{quintessence} and "phantom" with $w<-1$ \cite{phantom}, respectively.
Since current observational constraint on the equation of state of
dark energy lies in a relatively large range around the so-called
cosmological constant divide $w_X=-1$, it is still too early to rule
out any of the above candidates.

On the other hand, general relativity (GR) is very well examined in
the solar system, in observation of the period of the binary pulsar,
and in the early universe, via primordial nucleosynthesis. However,
no one has so far tested in the ultra-large length scales and low
curvatures characteristic of the Hubble radius today. Therefore, it
is a priori believable that Friedmann equation is modified in the
very far infrared, in such a way that the universe begins to
accelerate at late time. Freese and Lewis \cite{Freese2002}
construct so-called Cardassian universe models that incarnates this
hope. In Cardassian models \cite{Freese2002,Lazkoz,Freese03}, the
universe is flat and accelerating, and yet contains only matter
(baryonic or not) and radiation. But the usual Friedmann equation
governing the expansion of the universe is modified to be \be
\label{friedmanEQ1}
 H^2\equiv\left(\frac{\dot{a}}{a}\right)^2=\frac{8\pi G}{3}g(\rho),
\ee where $\rho$ consists only of matter and radiation, $H$ is the
Hubble "parameter" which is a function of time, $a$ is the scale
factor of the universe, and $G = 1/{m_{pl}^2}$ is the Newtonian
gravitational constant. Note that as required by  inflation scenario
and observations of CMB, the geometry of the universe is flat,
therefore, there are no curvature terms in the above equation.
Perhaps the most interesting feature of Cardassian models is that
although being matter dominated, they may be accelerating and may
still reconcile the indications for a flat universe ($\Omega_T=1$)
from CMB observations with clustering estimates that point
consistently to $\Omega_{m,0}=0.3$ with no need to invoke either a
new dark component or a curvature term. The expectation of
explaining cosmological observations without requiring new energy
sources or certainly worthy of investigation.

For any suitable Cardassian model, there exist at least three
requirements that should be satisfied. Firstly, the function
$g(\rho)$ should returns to the usual form $\rho$ at early epochs in
order to recover the thermal history of the standard cosmological
model and the scenario for the formation of large scale structure.
Secondly, $g(\rho)$ should takes a different form  at late times
$z\sim \mathcal{O}(1)$ in order to drive an accelerated expansion as
indicated by the observation of SNeIa \cite{SNeIa,SnST,SnLS}.
Finally, the classical solution of the expansion should be stable,
\ie,  the sound speed $c_s^2$ of classical perturbations of the
total cosmological fluid around homogeneous FRW solutions cannot be
negative.

For the original power-law Cardassian model $ \label{PowerLawModel}
g(\rho)=\rho +B\rho^n,$ where $B$ and $n<2/3$  are two constants
\cite{Freese2002}, the second term in the above equation behave like
an effective cosmic dark energy component that drive the universe
accelerate at late times and is negligible at early epochs. However,
the sound speed of cosmological fluid in this model is not
guaranteed to be positive. So this model should only be considered
as an effective description at scales where the sound speed is
positive \cite{Gondolo02}. The generalized power-law Cardassian
models, such as MP Cardassian model \cite{Freese03}, satisfies all
of above requirements, resorting to an additional parameter.

 In this paper, we investigate another kind of Cardassian models, dubbed Exponential
Cardassian models. In the next section, we construct two models that
embodies the elegant idea of Cardassian universe. In section 3, we
fit the parameters of the models to the type Ia supernovae
observations. The issue on the sound speed of the cosmological fluid
$c_s^2$ is addressed in section 4. And in the last section, a brief
discussion is included.

\section{Exponential Cardassian models}
\subsection{Model I: a simple exponential Cardassian model}
In this subsection we consider the following simple version of the
Exponential Cardassian model:
\begin{equation}\label{exponential Cardassian}
H^2=\frac{8\pi G}{3}g(\rho)=\frac{8\pi
G}{3}\rho\exp\left[\left(\frac{\rho_{card}}{\rho}\right)^n\right],
\end{equation}
where $\rho_{card}$ is a characteristic constant energy density and
$n$ is a dimensionless constant. For the late-time evolution of the
universe we neglect the contribution of radiation. The current
acceleration expansion of the universe requires that
\begin{equation}
3n\left(\frac{\rho_{card}}{\rho_0}\right)^n>1,
\end{equation}
where $\rho_0$ is the current value of energy density of matter
$\rho$ in the universe, which keeps conserved during the expansion
of the universe, \ie
\be\label{wuzhishouheng}\dot{\rho}+3H(\rho+p)=0. \ee Therefore, the
evolution of matter takes the ordinary manner
 \be\rho=\rho_0(1+z)^3.\ee

Obviously, at early times, $\rho$ is much larger than the
characteristic energy density $\rho_{card}$, $g(\rho)\rightarrow
\rho$, i.e. Eq.(\ref{exponential Cardassian}) recovers the standard
Friedmann equation. In terms of Eqs.(\ref{exponential Cardassian})
and (\ref{wuzhishouheng}), the effective pressure of total fluid
$p_T$ takes the following form:  \be\label{ptotal}
p_T=\rho\frac{\partial g(\rho)}{\partial \rho}-g(\rho).
 \ee

In the light of the observation that depend only on the scale
factor, in the late time regime of the universe filled with just
matter, Cardassian models undifferentiated from the effective dark
fluid models with the equation of state $p_X=w_X\rho_X$, where
$\rho_X=g(\rho)-\rho$ and $p_X=p_T$.

 Using the dynamics of $\rho$, we change the
Eq.(\ref{exponential Cardassian}) into the form
\begin{equation}\label{h2h02}
H^2=H_0^2\left[\Omega_{m,0}(1+z)^3+(1-\Omega_{m,0})f_{X}(z)\right],
\end{equation}
where the current value of Hubble parameter $H_0$ is often denoted
as $100\;h\; \mbox{km s}^{-1} \mbox{Mpc}^{-1}$ in which $h$ is a
dimensionless constant and
\begin{equation}\label{fxz1}
f_X(z)=\frac{\Omega_{m,0}(1+z)^3}{1-\Omega_{m,0}}\left\{\Omega_{m,0}^{-1}\exp\left[\left((1+z)^{-3n}-1\right)\ln\Omega_{m,0}^{-1}\right]-1\right\}.
\end{equation}
It is easy to check that $f_X(z=0)=1$. Therefore, in this model,
besides the density parameter $\Omega_{m,0}$, there is only one free
parameter $n$ for the effective dark fluid as in the original
power-law Cardassian model \cite{Freese2002}.

 Using Eq.(\ref{ptotal}), we can get the
effective parameter of equation of state $w_T$ for the total
cosmological fluid in this model
\begin{eqnarray}\label{PLptotalmatterradiation1}
w_T&\equiv&\frac{p_T(\rho)}{g(\rho)}=-n\frac{\rho_{card}^n}{\rho^{n}}=
n\ln(\Omega_{m,0})(1+z)^{-3n}.
\end{eqnarray}
However, the equation-of-state parameter of effective dark fluid
$w_X$ becomes
\begin{eqnarray}\label{PLptotalmatterradiation2}
w_X&\equiv
&\frac{p_X(\rho)}{\rho_X(\rho)}=\frac{p_T(\rho)}{g(\rho)-\rho}=
\frac{n\ln(\Omega_{m,0})(1+z)^{-3n}}{1-\exp\left[\ln\Omega_{m,0}(1+z)^{-3n}\right]}.
\end{eqnarray}

\begin{figure}
\centering
\includegraphics[width=8.5cm]{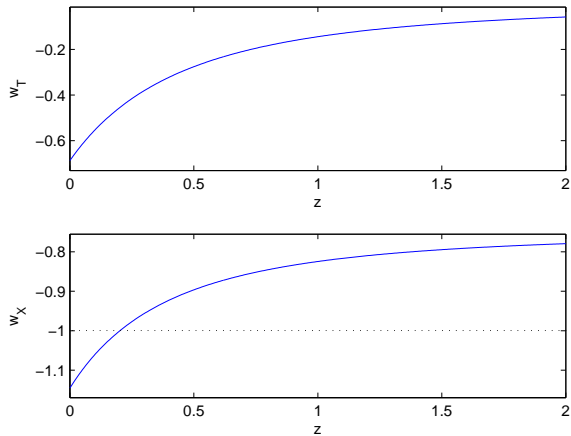}
\caption{The evolution of $w$ as a function of redshift $z$ in the
context of exponential Cardassian model where we have choose the
parameters $n=0.75$, $\Omega_{m,0}=0.40$. It is obvious that the
equation-of-state parameter of effective dark fluid $w_X$ (down
panel) cross the cosmological constant divide $w_{\Lambda}=-1$,
whereas that of total fluid is greater than $-1$ (up panel) up to
now.} \label{fig4}
\end{figure}

In Fig.\ref{fig4}, we show the evolutions of the equation-of-state
parameter for the total cosmological fluid $w_T$ and that for the
effective dark fluid. Clearly, $w_X$ (down panel) cross the
cosmological constant divide $w_{\Lambda}=-1$ recently, whereas
$w_T$ is always greater than $-1$ (up panel) up to now. However, we
note that both $w_T$ and $w_X$ will be less than $<-1$ and trend to
negative infinity with the expansion of the universe. Therefore, the
universe will suffer from the fate of so-call "big-rip"
 in the future \cite{bigrip}. Interestingly, $w_X$ is a negative
 constant in the original power-law Cardassian model
 \cite{Freese2002}. Therefore, this can lead to discrepancies
 between exponential and power-law models.

\subsection{Model II: a modified version of exponential Cardassian model}
We now consider the following modified version of the original
Cardassian model:
\begin{equation}\label{exponential_Cardassian_2}
g(\rho)=(\rho+\rho_{card})\exp\left[\left(\frac{q\rho_{card}}{\rho+\rho_{card}}\right)^n\right],
\end{equation}
where $\rho_{card}$ is a characteristic constant energy density and
$q$ and $n$ are two dimensionless positive constants. Obviously, at
early times, $\rho$ is much larger than the characteristic energy
density $\rho_{card}$, $g(\rho)\rightarrow \rho$, i.e.
Eq.(\ref{exponential_Cardassian_2}) recovers the standard Friedmann
equation. In terms of Eqs.(\ref{exponential_Cardassian_2}) and
(\ref{ptotal}), the current acceleration expansion of the universe
requires that
\begin{equation}
\sigma_0\left[1-3n \left(\frac{q}{1+\sigma_0}\right)^n\right]<2 ,
\end{equation}
where we define the dimensionless quantity
$\sigma\equiv\rho/\rho_{card}=\rho_{0}/\rho_{card}(1+z)^3\equiv
\sigma_0(1+z)^3$.According to the assumption that the universe is
flat, there is a relationship between the four parameters
$\Omega_{m,0}$, $\sigma_0$,
 $q$ and $n$ that
\begin{equation}\label{Paramrelation}
(1+\sigma_0)\exp\left[\left(\frac{q}{1+\sigma_0}\right)^n\right]=\sigma_0\Omega_{m,0}^{-1}.
\end{equation}

Using the dynamics of $\rho$ and Eq.(\ref{Paramrelation}), the
function $f_X(Z)$ in Eq.(\ref{h2h02}) that describe the evolution of
the effective dark fluid becomes
\begin{eqnarray}\label{fxz}
f_X(z)&=&\frac{\Omega_{m,0}}{1-\Omega_{m,0}}\left\{
\left[\sigma_0^{-1}+(1+z)^3\right]
\exp\left[\frac{-(1+\sigma_0)^n\ln[(1+\sigma_0^{-1})\Omega_{m,0}]}{\left[1+\sigma_0(1+z)^3\right]^{n}}\right]
\right.\nonumber\\
&&\hspace{2.0cm}\left.-(1+z)^3\right\}.
\end{eqnarray}

 Combining Eqs.(\ref{fxz}) and (\ref{Paramrelation}), it is easy to check that $f_X(z=0)=1$.
  Therefore, in this modified model, if we assume \emph{a priori}  value for
$\Omega_{m,0}$ , there is only two free parameters $q$ and $n$ for
the effective dark fluid as in the MP Cardassian model
\cite{Freese03}.

\begin{figure}
\centering
\includegraphics[width=8.5cm]{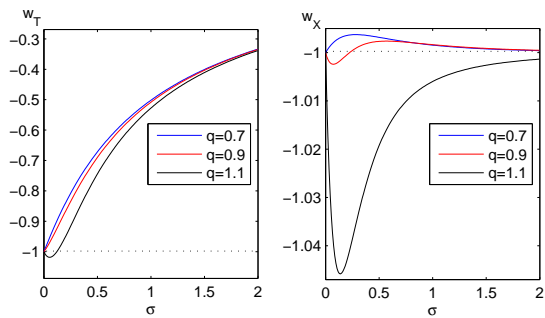}
\caption{The equation-of-state parameters $w_T$ and $w_X$ as two
functions of $\sigma$ in the modified version of exponential
Cardassian model for different values of parameter $q$, where we
have chosen $\Omega_{m,0}=0.30$, $n=5$.} \label{fig6}
\end{figure}

 Using Eq.(\ref{ptotal}), we can get the
effective parameter of equation of state $w_T$ for the total
cosmological fluid in this model
\begin{eqnarray}\label{PLptotalmatterradiation1.2}
w_T=-1+\frac{\sigma}{1+\sigma}\left[1+n\left(\frac{1+\sigma_0}{1+\sigma}\right)^n\ln[(1+\sigma_0^{-1})\Omega_{m,0}]\right],
\end{eqnarray}
and the equation-of-state parameter of effective dark fluid $w_X$
becomes
\begin{eqnarray}\label{PLptotalmatterradiation2.2}
w_X=
\frac{-1+{\sigma}n\left(\frac{1+\sigma_0}{1+\sigma}\right)^n\ln[(1+\sigma_0^{-1})\Omega_{m,0}]}
{1+\sigma\left[1-\exp\left(\left(\frac{1+\sigma_0}{1+\sigma}\right)^n\ln[(1+\sigma_0^{-1})\Omega_{m,0}]\right)\right]}.
\end{eqnarray}
In the limit of $\sigma\rightarrow 0$, both $w_T$ and $w_X$ trend to
$-1$, therefore, the expansion of the universe will speedup forever
and in the end the universe becomes de-Sitter universe
asymptotically. For the current epoch $\sigma(z=0)=\sigma_0$,
therefore, the current value of $w_X$ is determined by
\begin{equation}
w_X(z=0)=\frac{-1+n\sigma_0\ln[(1+\sigma_0^{-1})\Omega_{m,0}]}{(1+\sigma_0)(1-\Omega_{m,0})}.
\end{equation}
Obviously, the value of $w_X(z=0)$ can be greater or less than $-1$.
Furthermore, from Fig.\ref{fig6}, we can see the evolution of the
$w_X(z)$. It is interesting that, for some values of parameters,
$w_X$ can cross the cosmological constant divide $w_{\Lambda}=-1$ at
a time.

\section{Fit the model parameters to Supernovae data}

The Exponential Cardassian models predict a specific form of the
Hubble parameter $H(z)$ as a function of redshift $z$ in terms of
two parameters $\Omega_{m,0}$ and $n$. Using the relation between
$d_L (z)$ and the comoving distance $r(z)$ (where $z$ is the
redshift of light emission) \be \label{dlz1} d_L (z) = r(z) (1+z),
\ee and the light ray geodesic equation in a flat universe $c \; dt
= a(z) \; dr(z)$ where $a(z)$ is the scale factor.

 In general, the approach
towards determining the expansion history $H(z)$ is to assume an
arbitrary ansatz for $H(z)$ which is not necessarily physically
motivated  but is specially designed to give a good fit to the data
for $d_L (z)$. Given a particular cosmological model for $H(z; a_1,
... ,a_n)$ where $a_1, ...,a_n$ are model parameters, the maximum
likelihood technique can be used to determine the best fit values of
parameters as well as the goodness of the fit of the model to the
data. The technique can be summarized as follows: The observational
data consist of $N$ apparent magnitudes $m_i (z_i)$ and redshifts
$z_i$ with their corresponding errors $\sigma_{m_i}$ and
$\sigma_{z_i}$. These errors are assumed to be gaussian and
uncorrelated. Each apparent magnitude $m_i$ is related to the
corresponding luminosity distance $d_L$ by \be \label{mz1} m(z)=M +
5 \; \log_{10} \left[{{d_L (z)}\over {\mbox{Mpc}}}\right] + 25, \ee
where $M$ is the absolute magnitude. For the distant SNeIa, one can
directly observe their apparent magnitude m and redshift z, because
the absolute magnitude $M$ of them is assumed to be constant,\ie,
the supernovae are standard candles. Obviously, the luminosity
distance $d_L (z)$ is the `meeting point' between the observed
apparent magnitude $m(z)$ and the theoretical prediction $H(z)$.
Usually, one define distance modulus $\mu(z)\equiv m(z)-M$ and
express it in terms of the dimensionless `Hubble-constant free'
luminosity distance $D_L$ defined by$D_L (z) = {{H_0 d_L (z)}/ c}$
as
\begin{equation}
\label{mz2}
 \mu(z)=5\; \log_{10}(D_L (z))+\mu_0,
\end{equation}
where the zero offset $\mu_0$ depends on $H_0$ (or $h$) as
 \begin{equation}
\label{bm1}
\mu_0 =5\; \log_{10} \left({{cH_0^{-1}}\over
{\mbox{Mpc}}}\right) + 25=-5\log_{10}h+42.38.
\end{equation}
The theoretically predicted value $D_L^{th} (z)$ in the context of a
given model $H(z;a_1,...,a_n)$ can be described by
\cite{Starobinsky,Huterer,Chiba}
 \begin{equation}
 \label{dth1}
 D_L^{th} (z) = (1+z)
\int_0^z dz' \; {{H_0}\over {H(z';a_1,...a_n)}}.
\end{equation}
Therefore, the best fit values for the parameters ($\Omega_{m,0},n$)
of model I are found by minimizing the quantity
 \begin{equation}
 \label{chi2def}
 \chi^2 (\Omega_{m,0},n)=\sum_{i=1}^N
 \frac{\left[\mu^{{obs}}(z_i)-5\log_{10}D_L^{{th}}(z_i;\Omega_{m,0},n)-\mu_0\right]^2}{\sigma_i^2}.
\end{equation}
Since the nuisance parameter $\mu_0$ is model-independent, its value
from a specific good fit can be used as consistency test of the data
\cite{Choudhury} and one can choose \emph{a priori} value of it
(equivalently, the value of dimensionless Hubble parameter $h$) or
marginalize over it thus obtaining
\begin{equation}
\widetilde{\chi}^2(\Omega_{m,0},n)=A(\Omega_{m,0},n)
-\frac{B(\Omega_{m,0},n)^2}{C}+\ln\left(\frac{C}{2\pi}\right),
\end{equation}
where
\begin{equation}\label{A}
   A(\Omega_{m,0},n)=\sum_{i=1}^N
 \frac{\left[\mu^{{obs}}(z_i)-5\log_{10}D_L^{{th}}(z_i;\Omega_{m,0},n)\right]^2}{\sigma_i^2},
\end{equation}
\begin{equation}\label{B}
       B(\Omega_{m,0},n)=\sum_{i=1}^N
 \frac{\left[\mu^{{obs}}(z_i)-5\log_{10}D_L^{{th}}(z_i;\Omega_{m,0},n)\right]}{\sigma_i^2},
\end{equation}
and
\begin{equation}\label{C}
   C=\sum_{i=1}^N
 \frac{1}{\sigma_i^2}.
\end{equation}
In the latter approach, instead of minimizing
${\chi}^2(\Omega_{m,0},n)$, one can minimize
$\widetilde{\chi}^2(\Omega_{m,0},n)$ which is independent of
$\mu_0$.

We now apply the above described maximum likelihood method using
Gold dataset which is one of the reliable published data set
consisting of 157 SNeIa ($N=157$) \cite{SnST}.

\begin{figure}
\centering
\includegraphics[width=6.5cm]{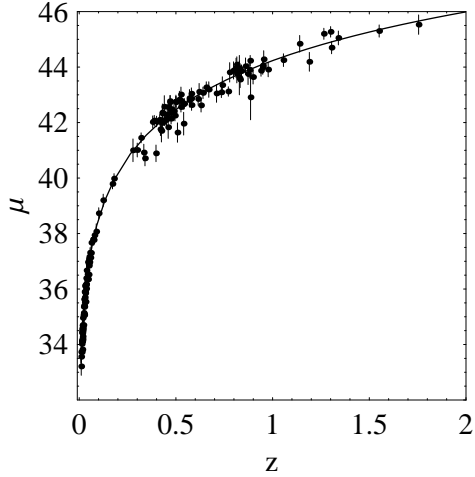}
\caption{The observed 157 Gold SNeIa distance modulus along with the
theoretically predicted curves in  model I (solid line), where we
take \emph{a priori} that current dimensionless Hubble parameter
$h=0.65$.} \label{fig3}
\end{figure}

 In Fig. \ref{fig3}, we show a comparison of the observed 157 SNeIa
distance moduli along with the theoretically predicted curves in
model I (continuous line). It is obvious that the exponential
Cardassian model provide a good fit to the observational data.

\begin{figure}
\centering
\includegraphics[width=9.5cm]{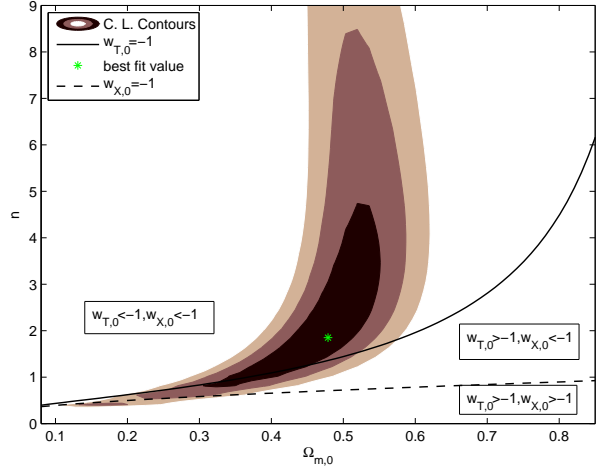}
\caption{The 68.3\%, 95.4\% and 99.7\% confidence contours of
parameters $\Omega_{m,0}$ and $n$ using the Gold SNeIa dataset and
marginalizing over the model-independent parameter $\mu_0$.}
\label{fig1.t2}
\end{figure}

In Fig.\ref{fig1.t2},  contours with 68.3\%, 95.4\% and 99.7\%
confidence level are plotted, in which we take a marginalization
over the model-independent parameter $\mu_0$. The best fit as showed
in the figure corresponds to $\Omega_{m,0}=0.48$ and $n=1.84$, and
the minimum value of $\chi^2=173.51$. Clearly, the allowed ranges of
the parameters $\Omega_{m,0}$ and $n$ favor that there exists an
effective phantom energy in the universe.

As for model II, using the above marginalization method, we find the
minimum value of $\chi^2(\Omega_{m,0},\sigma_0,n)$ is $173.08$ and
the corresponding best-fit value of parameters are
$\Omega_{m,0}=0.308$, $\sigma_0=0.708$ and $1/n=0.0142$. If  we
assume prior that $\Omega_{m,0}=0.3$ as indicated by the observation
about mass function of galaxies, the minimum value of
$\chi^2(\sigma_0,n)$ is $173.178$ corresponding to $\sigma_0=0.694$
and $n^{-1}=0.0123$.

\begin{figure}
\centering
\includegraphics[width=9.5cm]{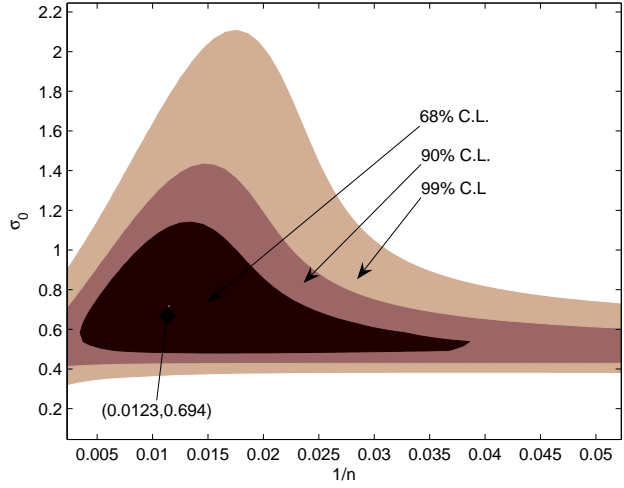}
\caption{The 68\%, 90\% and 99\% confidence level contours of
parameters $\Omega_{m,0}$ and $n$ using the Gold SNeIa dataset and
marginalizing over the model-independent parameter $\mu_0$ with a
prior $\Omega_{m,0}=0.3$.} \label{fig1.t3}
\end{figure}

\section{On the sound speed $c_s^2$}
In order to guarantee the classical solution of the expansion is
stable, the sound speed $c_s^2$ of classical perturbations of the
total cosmological fluid around homogeneous FRW solutions should
always be greater than zero. If the expansion of the universe is
adiabatic, the sound speed of total cosmological fluid can be
represented by $c_s^2=\delta p_T/\delta \rho_T$. For model I,
\begin{equation}
c_s^2=-n\ln(\Omega_{m,0}^{-1})(1+z)^{-3n}
\frac{n\ln(\Omega_{m,0}^{-1})(1+z)^{-3n}+n-1}{n\ln(\Omega_{m,0}^{-1})(1+z)^{-3n}-1}.
\end{equation}
In order to assure that $c_s^2>0$ at any time, we should have
\begin{equation}\label{cond1}
n\ln(\Omega_{m,0}^{-1})(1+z)^{-3n}+n-1>0
\end{equation}
 and
 \begin{equation}\label{cond2}
    n\ln(\Omega_{m,0}^{-1})(1+z)^{-3n}-1<0.
 \end{equation}
From condition (\ref{cond1}), $n$ should be greater  than $1$; but
condition (\ref{cond2}) can not be satisfied for every value of
$z\in(-1,+\infty)$. Therefore, model I should only be considered as
an effective description for the cosmic expansion. However, we note
that if we only require that $c_s^2>0$ for $z>0$, \ie, classical
solution of the expansion is stable in the past, then we should let
$-1<w_{T,0}<n-1$ which is clearly satisfiable.

As for model II, the sound speed can be denoted as
\begin{equation}\label{1}
    c_s^2=-\frac{n\sigma}{q}\left(\frac{q}{1+\sigma}\right)^{n+1}\frac{n\left(\frac{q}{1+\sigma}\right)^n+n-1}{n\left(\frac{q}{1+\sigma}\right)^n-1}.
\end{equation}
It is easy to find that $c_s^2$ is always positive if the parameters
satisfy the condition $q^n<n^{-1}<1$.

\section{Discussion}
\label{} In above sections, we have investigated a new kind of
Cardassian  models of the universe,dubbed Exponential Cardassian
universe. Contrary to the original power-law Cardassian model
\cite{Freese2002}, the equation-of-state parameter $w$ of effective
dark fluid is dependent on time that can cross the cosmological
constat divide $w_{\Lambda}=-1$ from $w_X>-1$ to $w_X<-1$ as the
observations mildly indicate.

However, it is worth noting that these two models are available at
much large scales where the matter density $\rho_m \sim \rho_c$ and
evolve in the light of $\propto (1+z)^{-3}$. As discussed by Gondolo
and Freese \cite{Gondolo02}, the gradient of the effective
Cardassian pressure $\nabla p_{card}$ should be able to neglected
compared to that of the ordinary pressure $\nabla p_{m}$ at the
scales of galaxies and galaxy clusters where the matter density is
much larger than cosmic average density. In a typical galaxy where
$\rho_m \sim 100 \rho_c \sim 100 \rho_{card}$ and $p_m=\rho_m
\Sigma^2$ with velocity dispersion $\Sigma=300\; \mbox{km/s}$, if we
assume that $\nabla p_{card}\ll\nabla p_{m}$, we should require the
parameter $n$ in both models considered above are greater than $3$.
It is remarkable that this value of $n$ is compatible with the data
of SNeIa observations. Therefore, these two models are both
available on galactic scales.

Observationally, the exponential Cardassian models are compatible
with the data of SNeIa. The further work that should do is to check
it out by other cosmological and astrophysical observations such as
CMB and large scale structures (LSS) as many works (for example,
\cite{koivisto,constraints}) for power-law Cardassian models, which
is out of the scope of this paper. However, it is worth noting that
Koivisto \emph{et al} \cite{koivisto} pointed out that in the MP
Cardassian model, the late integrated Sachs-Wolf effect is typically
very large duo to the suppression of fluctuations in Cardassian
fluid at late times, and is then not compatible with the
observations of CMB unless the parameters is very close to the
$\Lambda$CDM model. In the exponential Cardassian models, we assume
that Cardassian fluctuations is induced by cold dark matter (i.e.,
case III in Ref. \cite{koivisto}) and fluctuations in the cold dark
matter and in the Cardassian fluid are related adiabatically
\begin{equation}
\delta=\left(\frac{d\log
\rho_T}{d\log\rho}\right)^{-1}\delta_T=\frac{\delta_T}{1+w_T},
\end{equation}
where $\delta$ and $\delta_{T}$ denote the fluctuations in cold dark
matter and Cardassian fluid, respectively. According to the
expressions of $w_T$, Eq.(\ref{PLptotalmatterradiation1}) for model
I and Eq. (\ref{PLptotalmatterradiation1.2}) for model II, it is not
hard to find that the Cardassian fluctuations are both not heavily
suppressed as those in MP Cardassian model  which $\delta\sim
a^{3q}\delta_{T}$ at late times \cite{koivisto}. Therefore, in the
exponential Cardassian models the above problem is alleviated to
some extent.

 Theoretically, these models, in fact as well as
power-law Cardassian models, are phenomenologically  described in a
fluid approach, therefore, we need further works to find a simple
fundamental theory which can derive the form of $g(\rho)$ presented
above.

\section*{Acknowledgements}
This work is supported by National Natural Science Foundation of
China under Grant No. 10473007 and No. 10503002.

\end{document}